\documentclass[sigconf, screen]{acmart}
\AtBeginDocument{%
	\providecommand\BibTeX{{%
			\normalfont B\kern-0.5em{\scshape i\kern-0.25em b}\kern-0.8em\TeX}}}

\copyrightyear{2024}
\acmYear{2024}
\setcopyright{rightsretained}
\acmConference[DIS '24]{Designing Interactive Systems Conference}{July 1--5, 2024}{IT University of Copenhagen, Denmark}
\acmBooktitle{Designing Interactive Systems Conference (DIS '24), July 1--5, 2024, IT University of Copenhagen, Denmark}\acmDOI{10.1145/3643834.3660730}
\acmISBN{979-8-4007-0583-0/24/07}

\usepackage{subcaption}
\usepackage{graphicx}
\begin{document}
	
	\title{Enhancing Reentry Support Programs Through Digital Literacy Integration}
	
	\author{Aakash Gautam}
	\affiliation{%
		\institution{University of Pittsburgh}
		\country{Pittsburgh, PA, USA}
	}
	\email{aakash@pitt.edu}
	
	\author{Khushboo Gandhi}
	\affiliation{%
		\institution{San Francisco State University}
		\country{San Francisco, CA, USA}
	}
	\email{kgandhi1@mail.sfsu.edu}
	
	\author{Jessica Eileen Sendejo}
	\affiliation{%
		\institution{San Francisco State University}
		\country{San Francisco, CA, USA}
	}
	\email{jsendejo1@mail.sfsu.edu}
	
	
	\begin{abstract}
		
		
		Challenges faced by formerly incarcerated individuals in the United States raise questions about our society's ability to truly provide second chances. This paper presents the outcomes of our ongoing collaboration with a non-profit organization dedicated to reentry support. We highlight the multifaceted challenges individuals face during their reentry journey, including support programs that prioritize supervision over service, unresponsive support systems, limited access to resources, financial struggles exacerbated by restricted employment opportunities, and technological barriers. In the face of such complex social challenges, our work aims to facilitate our partner organization's ongoing efforts to promote digital literacy through a web application that is integrated into their existing processes. 
		We share initial feedback from the stakeholders, draw out four implications: supporting continuity of care, promoting reflection through slow technology, building in flexibility, and reconfiguring toward existing infrastructure, and conclude with a reflection on our role as partners on the side.
	\end{abstract}

	

	\maketitle

	\section{Introduction}

	The United States is a ``carceral state'', with an estimated 1.9 million people in prison.
	Each year upwards of 600,000 people are released from state and federal prisons \cite{incarcerationReentry, sawyer2023mass}. 
	Formerly incarcerated individuals, who we shall henceforth call ``returning community members''\footnote{In literature, we find the use of the term ``returning citizens'' (e.g., \cite{Ogbonnaya-Ogburu_2019, ortiz2019system}), but many of the participants we worked with are not US citizens. Around 15\% of people incarcerated in the US are not US citizens \cite{carson2021federal}. Acknowledging this, our partner organization uses the term ``returning community members'', which we adopt in this paper.}, face a multitude of challenges during their reintegration journey. 
	These challenges include limited access to resources, educational disparities, stigmatization, and limited employment prospects. 
	Structural support in overcoming these challenges is mostly limited, leading to a high rate of recidivism, perpetuating a cycle of re-incarceration \cite{miller2014devolving, ortiz2019system, smith2010african}. 
	All of these raise questions about whether our existing social system can truly provide returning community members the second chance they have been promised. 
	
	Our work is situated in this context. 
	For the past one and a half years, we have been collaborating with a non-profit organization in San Francisco (California). 
	The city has a history of a high incarceration rate \cite{center1994race}. 
	In addition, there is staggering racial disparity; in California, Black and Latinx individuals are  9.5 and 2.0 times more likely to be incarcerated than White Americans \cite{nellis2016color, prisonPolicyInitiative}. 
	Our partner organization, Project Rebound, is part of a public university. 
	Project Rebound helps students obtain a formal degree in one of the many public educational institutions in and around San Francisco. 
	In addition, Project Rebound provides support through informal literacy programs and advising.


	In this paper, we reflect on our ongoing journey with Project Rebound and, through it, we make four-fold contributions to human-computer interaction (HCI) scholarship.
	First, we highlight the multifaceted and enduring challenges returning community members face during reentry, even when enrolled in support programs like those provided by our partner organization. 
	These include restrictive transitional housing, unresponsive parole systems, financial struggles, barriers to employment, and difficulties navigating an increasingly complex digital ecosystem.
	Second, we present our approach for enhancing our partner organization's efforts in supporting digital literacy. 
	Our iterative, collaborative design process with the partner organization resulted in a web application that aligns with and enhances their existing practices.
	Third, we reflect on lessons learned for designing digital literacy support, especially for marginalized groups.
	This includes supporting continuity of care by incorporating scaffolds for interdependence and collaboration, promoting reflection through intentionally slow design, building in flexibility to accommodate differences in users' contexts and digital skills, and reconfiguring design to leverage existing infrastructures.
	Finally, we articulate and argue for the value of being a \emph{partner on the side}, leveraging our technical expertise to facilitate, rather than drive, community-based efforts. 
	This role enables supporting organizational ownership and centers continuity in the collaboration.
	

	\section{Literature Review}
	
	\subsection{Background: Mass Incarceration and Prisoner Reentry Programs in the United States}
	\label{Sec:Background}
	
	In the early 1970s, Richard Nixon's government initiated the War On Drugs which resulted in mass incarceration.
	This surge in incarceration rates marked a shift away from rehabilitation-focused programming towards a ``warehousing model of incarceration'' \cite{simon2010beyond}, wherein prisons primarily served as instruments of regulation and punishment rather than addressing the root causes of crime \cite{ortiz2019system, wacquant2009punishing}. 
	At the same time, the growth of neoliberal ideologies in the US led to a dissolution of the welfare state, and with it, privatized prisons---aiming to profit from mass incarceration---grew exponentially \cite{davis2000masked, morris2007government, fulcher2011hustle, widra2020since}, further limiting opportunities for rehabilitation \cite{mauer2012bureau}.
	
	The War On Drugs skewed heavily against Black Americans, causing the ``Blackening of prisons'' \cite{miller2014devolving, michelle2010new}. 
	The prison demographic changed ``from over 70\% White in the 1950s to  70\% Black and Latino by 1989'' \cite[pp. 485]{ortiz2019system} (see also \cite{wacquant2001deadly, pettit2004mass}). 
	This racial disparity persists today, with Black Americans incarcerated at five times the rate of white Americans, and in some states, including our own, the ratios exceed 9:1 \cite{nellis2016color}. 
	Moreover, poor people were, and continue to be, disproportionately incarcerated \cite{rabuy2015prisons, clear2009imprisoning, pettit2004mass, akee2019race, bronson2015disabilities}. 
	The expansion of privatized prisons combined with these discriminating practices, underscore a focus on control over historically marginalized groups rather than rehabilitation \cite{miller2014devolving, wacquant2009punishing, ortiz2019system}.
	
	Prisoner reentry programs have not escaped the influence of neoliberalism, leading critics to label them a ``prison reentry industry'' (e.g., \cite{davis2000masked, thompkins2010expanding, smith2010african}).
	Indeed, it is a burgeoning industry, with one in 48 U.S. adults under some form of correctional supervision in 2021 \cite{carson2022correctional}.
	
	The problem is structural and, from our perspective, vicious. 
	The ``school-to-prison'' pipeline \cite{pettit2004mass} funnels children out of school to the juvenile and criminal justice systems. 
	It has hindered the educational attainments of returning community members.
	A quarter of these individuals have less than a high school diploma and only 4\% hold a college degree \cite{couloute2018getting}. 
	Rehabilitation programs within prisons are severely limited \cite{lee2019halfway, wakefield2010incarceration}. 
	Programs outside the prison also prioritize monitoring and surveillance over rehabilitation as seen in the far fewer and limited literacy or mental health programs (e.g., \cite{binswanger2012return, smith2010african, begun2016mental, harding2014making}).
	Additional restrictions imposed by the criminal justice system \cite{halushka2019runaround, trimbur2009me} and stigma in society further hinder returning community members' prospects for employment and other opportunities \cite{michelle2010new}.

	\subsection{Digital Literacy and Reentry}
	
	
	Digital literacy has become an essential skill to accomplish a range of tasks such as finding housing, education and learning, and accessing various public and social services \cite{van2014digital, lee2023collective, biehl2022can}.
	The need to master several digital skills to accomplish seemingly ``simple'' everyday tasks is growing \cite{allmann2021rethinking, van2012evolution}.
	Add to this the biases and inequalities enforced through the seemingly invisible but widely pervasive algorithmic decision-making systems that have been shown to disadvantage people of color as well as those with lower income and educational attainments (e.g., \cite{eubanks2018automating, benjamin2019race, noble2018algorithms, lageson2020digital, courtois2016little}).
	Thus, digital inclusion programs are imperative for reentry support \cite{zivanai2022digital, Ogbonnaya-Ogburu_2019, Reisdorf_2022}.
	
	A pertinent example of the effects of limited digital literacy in reentry efforts can be seen in seeking income-generating opportunities.
	Stable income, primarily through employment or entrepreneurship, is well-known to be pivotal for successful reentry \cite{visher2008employment, freeman2003can, henry2007ban}.
	The significance is resonated in initiatives such as the Biden-Harris administration's ``Incarceration to Employment'' strategic program\footnote{\url{https://www.whitehouse.gov/wp-content/uploads/2022/04/Incarceration-to-Employment-Strategy.pdf}}.
	However, the lack of digital literacy is a formidable barrier to employment opportunities. 
	Online resources and other digital infrastructures, such as LinkedIn, play a central role in job search, placing returning community members at a disadvantage \cite{Ogbonnaya-Ogburu_2019}. 
	A lack of understanding of digital rights that protect individuals from biases (for example, Federal and State laws prohibiting background inquiries such as with the ``ban-the-box'' policy or the Second Chance Ordinance \cite{goode2011internet, henry2007ban}) adds another layer of complexity. 
	
	HCI research on supporting the digital literacy of those returning from incarceration remains limited, with the noteworthy exception of \citet{Ogbonnaya-Ogburu_2019}.
	Their study builds upon a theoretical lens for ``digital rehabilitation'' \cite{Reisdorf_2018}, which advocates for comprehensive digital literacy programs during and after incarceration (see also \cite{Reisdorf_2021, Reisdorf_2022}).
	Building on their digital literacy sessions with a group of returning community members in Detroit, \citet{Ogbonnaya-Ogburu_2019} emphasize contextualizing digital literacy within larger goals such as job search or entrepreneurship and posit the importance of flexible approaches that attend to differences, including in digital skills and modalities of computing and technology.   
	On this note, \citet{grierson2022design} explored considerations for digital services to support returning community members in Australia. 
	While not particularly focused on digital literacy, their work highlights the importance of supporting independence and control, as well as attending to the diverse needs of the returning community members. 
	The work briefly discusses the need to support digital literacy, arguing for literacy programs while people are incarcerated, a space where several challenges abound \cite{taugerbeck2019digital}. 
	We concur with these recommendations and through this work, we add practical guidance on \emph{how} to design such approaches considering the contextual complexities associated with reentry.

	
	Moreover, efforts to promote digital literacy have often involved time-limited classes and sessions, potentially leading to limited access beyond it, which can have detrimental consequences \cite{gonzales2016contemporary}. 
	Recognizing the necessity for sustained support, our approach involves collaborating with an organization with longstanding experience in the field to support their ongoing efforts.
	We reflect on this experience to illuminate the particular entailment that needs to be considered in such partnerships. 
	With the increasing number of reentry supporting initiatives by organizations such as Project Rebound, Operation Restoration\footnote{\url{https://www.or-nola.org/}}, From Prison Cells to PhD\footnote{\url{https://www.fromprisoncellstophd.org/}}, and Reimagine Reentry\footnote{\url{https://www.letsreimaginereentry.org/}}, it behooves us to examine how we can design for--\emph{and} with--them such that they can foster digital literacy among returning community members.

	\subsection{HCI For and With Community Partners}
	
	HCI research has increasingly embraced community-based approaches to technology design and research (e.g., \cite{harrington2019deconstructing, moulder2014hci, parker2012health, oleary2022community, karusala2023care}).  
	In this space, scholars have called attention to equitable and fair processes in defining problems, methods, and measures of success \cite{vandenberghe2016designing, pierre2021getting, asad2015illegitimate, gitau2009fair} while also arguing for care in building and sustaining relationships \cite{biorn2018building, cooper2022systematic, tseng2022care}. 
	HCI Scholarship has also taken a more reflexive approach, highlighting challenges around navigating power differentials, promoting mutual learning, and attending to the dynamics and differences within a community \cite{dombrowski2016social, cooper2022systematic, fox2014community, disalvo2010hci, jiang2022understanding}. 
	
	Non-profit organizations have typically been the entities enabling collaboration with communities, facilitating access to difficult-to-reach populations (e.g., \cite{dillahunt2022village, gautam2020crafting, Ogbonnaya-Ogburu_2019}), and functioning as experts on the ground (e.g., \cite{grierson2022design, moulder2014hci, karusala2023care}). 
	Scholars have argued to position non-profits and community members not as passive beneficiaries but as partners in the engagement \cite{le2015strangers, gitau2009fair}.
	However, scholars also critique the trend of unequal partnerships, marked by labor, affective, and epistemic burdens on non-profits and community members \cite{cooper2022systematic, pierre2021getting, harrington2019deconstructing}. 
	So, while community-based engagements provide a way for democratizing knowledge production and meaningful social change, the majority of research remains extractive and academically self-serving \cite{pierre2021getting, harrington2019deconstructing, liang2021embracing, pal2017chi4good}. 
	
	In response to these issues, aligning with \citet{pierre2021getting}, we advocate for de-centering research-driven agenda and prioritizing existing community efforts and goals. 
	In arguing to support ongoing efforts rather than initiating new ones, our approach builds on assets-based design that underscores the importance of recognizing and leveraging existing strengths and resources within the community \cite{pei2019we, dickinson2019cavalry, wong2021reflections}. 
	\citet{wong2021reflections} posit the need to embed trust-building elements, facilitate the formation of an interdependent collective, and make moves towards incremental transformations in assets-based design work in communities. 
	They further argue for positioning technology as an intermediary for slow incremental transformations \cite{wong2021reflections}.
	We echo these values through our work. 
	Particularly, we highlight our role as a partner on the side through which we sought to integrate the outcomes of our engagement into the partner organization's existing processes. This helped ensure our contributions aligned with and enhanced the community partner's ongoing efforts.


	\section{Methodology}

Our ongoing one-and-a-half-year-long engagement began with initial conversations with Project Rebound, following an approach recommended in community-based work \cite{emejulu2022critical}.
Through this dialogue, we learned about Project Rebound's efforts, including the challenges they faced in supporting digital literacy. 
We saw this as an avenue for collaboration. 
However, we were unaware of the realities surrounding reentry. 
Thus, we undertook an incremental study, first learning from the partner organization and then engaging with a group of returning community members.
With each iteration, we gained deeper insights into the technical and social challenges surrounding reentry. 

\begin{table*}[]
	\caption{Participants in the two phases of the study.}
	\begin{tabular}{|l|l|l|l|l|l|}
		\hline
		\multicolumn{1}{|c|}{\textbf{Pseudonames}} & \multicolumn{1}{c|}{\textbf{Age}} & \multicolumn{1}{c|}{\textbf{Gender}} & \multicolumn{1}{c|}{\textbf{\begin{tabular}[c]{@{}c@{}}When were they\\ last in the system?\end{tabular}}} & \multicolumn{1}{c|}{\textbf{\begin{tabular}[c]{@{}c@{}}Self-reported confidence\\in their ability to learn \\ technology (out of 5)\end{tabular}}} & \multicolumn{1}{c|}{\textbf{\begin{tabular}[c]{@{}c@{}}Participated \\ in Study\end{tabular}}} \\ \hline
		OW                                        & 53                                & M                                    & 2018                                                                                                       & 4                                                                                                               & Initial Study                                                                                   \\ \hline
		JD                                        & 49                                & F                                    & 2021                                                                                                       & 3                                                                                                               & Initial Study                                                                                   \\ \hline
		JR                                        & 48                                & M                                    & 2003                                                                                                       & 2                                                                                                               & Initial Study                                                                                   \\ \hline
		LR                                        &  38                                 & F                                    & 2022                                                                                                       & 3                                                                                                                & Initial Study                                                                                   \\ \hline
		EO                                        & 42                                & M                                    & 2021                                                                                                       & 5                                                                                                               & Both Studies                                                                                   \\ \hline
		JT                                        & 52                                & M                                    & 2020                                                                                                       & 5                                                                                                               & Both Studies                                                                                  \\ \hline
		AP                                        & 43                                & M                                    & 2018                                                                                                       & 5                                                                                                               & Both Studies                                                                                   \\ \hline
		EH                                        & 62                                & M                                    & 2018                                                                                                       & 4                                                                                                               & Both Studies                                                                                 \\ \hline
		GJ (Staff)                                & 72                                & M                                    & 1998                                                                                                        & -                                                                                                               & Both Studies                                                                                 \\ \hline
		JN (Staff)
		& 51                                 & M                                    & 2001                                                                                                          & -                                                                                                               & Both Studies                                                                               \\ \hline
		DR (Staff)                                & 58                                 & F                                    & N/A                                                                                                          & -                                                                                                               & Both Studies                                                                                          \\ \hline
		MN (Staff)                                & 27                                 & M                                    & N/A                                                                                                          & -                                                                                                               & Evaluation Study                                                                               \\ \hline
		CY (Staff)                                       & 57                                 & M                                    & 2021                                                                                                          & -                                                                                                               & Evaluation Study                                                                               \\ \hline
		LX                                        & 63                                 & M                                    & 2022                                                                                                          & -                                                                                                               & Evaluation Study                                                                               \\ \hline
	\end{tabular}
	\label{tab:participants}
\end{table*}

\subsection{Reflexivity and Positionality}
Before we discuss the methodology, we reflect on our positionality that shapes the analysis and the overall engagement. 
We are a group of three academic researchers, who were back then associated with the same institution as our partner organization.
We believe that as a society, we are only as strong as our most vulnerable members.
Our work seeks to design socio-technical systems that support communities to have power in enacting the future they desire. 
Our expertise is technical and we offer it as part of the collaboration.
We acknowledge that the technical expertise helped enable the initial collaboration and has inherently shaped this work.

We come from different socio-economic backgrounds with diverse ethnic and gender identities; two of the authors are migrants to the United States and knew of different approaches to reentry back in their home countries.   
Neither we nor our immediate family members have undergone a reentry journey. 
The interviews and analysis involved moments of overwhelming emotional load. 
We acknowledge the privilege we hold in being able to detach ourselves from this work when necessary, a privilege that may not be readily available to the participants.

We have sought to build a relationship of trust with Project Rebound as we see this work as part of a long-term collaboration. 
This entailed striving for transparency and mutuality. 
For instance, we shared intermediary observations and findings after the analysis (see Section \ref{sec:researchApproach}). 
We also shared and discussed this draft with the staff in the organization so that we are transparent about what we are saying about their efforts \cite{alcoff1991problem}.   

\subsection{Research Approach}
\label{sec:researchApproach}
We initiated our research after we received IRB approval from our institution. 
Our incremental study can be broadly categorized into two phases: an initial inquiry and a subsequent prototyping and technology evaluation study.

\subsubsection{Initial Inquiry Phase}

We interviewed three key staff members at Project Rebound. 
The questions focused on their activities, the challenges they faced in their work, and the vision they have for the organization. 
Subsequently, we held a focus group session to share our reflections on the interviews, and discussed a course of action.

Following this, we invited eight returning community members associated with Project Rebound to participate in a three-session focus group discussion (see Table \ref{tab:participants}).
In the first session, we discussed the challenges faced by these individuals during the reentry process and their aspirations for the future.
The second session focused on the technology they use in their daily life, the barriers they encounter, and the approaches they employ to overcome those barriers. 
The third session involved envisioning the next steps in their journey, both within and beyond the university. 
Each of the three sessions lasted between one to one and a half hours.

\subsubsection{Prototyping and Technology Evaluation Phase}
In the second phase, we engaged in an iterative design process with the partner organization. 
We were wary of burdening them through constant iterations---which is central in design--so, following \citet{dourish2020being}, we sought to find a balance by engaging them at points where we could show meaningful progress. 

We analyzed the interviews and focus group discussions from the first phase (see Section \ref{sec:dataCollectionAnalysis}). 
We presented key insights gained from the analysis which helped convey progress.
It also helped scaffold the brainstorming sessions.
In the discussion, priority was placed on supporting digital literacy. 
For example, the idea of assisting with resume-building emerged through this discussion.
Following this session, only the researchers engaged in thinking through the brainstormed ideas and narrowing them down to a few priority areas. 
Once we had converged on a few possible ways to support digital literacy (e.g., one-on-one mentoring, web application, call-based support), we presented those to Project Rebound staff to gather their feedback. 
Their input helped in refining the design ideas, leading us to prioritize a web application.

We began designing the web application. 
Following typical interaction design processes \cite{PreeceRogersSharp15}, we iterated internally until we had a high-level prototype. 
At this point, we ran usability testing sessions with both the staff and returning community members, asking them to complete tasks using the prototype while thinking aloud. 
We observed their interactions, noted any challenges or confusion, and gathered feedback on the overall user experience. 
Based on this, we arrived at a design used in the final web application and evaluation study, which was presented to a group of five staff members and five returning community members.  

This approach enabled us to ask less of our partners' time and protect them from some of the affective demands of sharing early, unfinished prototypes. 
It still provided key moments for gaining their input, ensuring the final designs matched their needs and perspectives. 
The insights gathered through these iterations informed the development of our web application, which now serves as a tangible outcome of our collaboration.


\subsection{Data Collection and Analysis}
\label{sec:dataCollectionAnalysis}

The returning community members who participated in our studies were associated with Project Rebound. 
The demographic distribution of the participating returning community members closely mirrored the broader population of returning community members. For instance, to illustrate, approximately 13\% of returning community members are female \cite{miller2021female}. 
To build in some level of reciprocity, each participant received compensation of \$25 per session. 
This amounted to \$75 for participating in the initial study and \$25 for the evaluation study.

We took field notes during conversations with the staff and returning community members.
In addition, we recorded audio during the staff interviews and focus group discussions, as well as the group sessions with the returning community members. 
This resulted in upwards of 13 hours of audio recordings.

The recorded interviews were first transcribed. 
We then conducted open coding by following the transcripts closely. In this process, we created summaries line by line \cite{saldana2015coding}. 
Through multiple rounds of discussion and iterations on the codes, we merged 962 low-level codes to eventually arrive at 31 higher-level codes.  
Some of the higher-level codes include ``transitional housing challenges'', ``lack of trust'', ``creating a sense of safety'', and ``privacy concerns with technology''.  
After finalizing the higher-level codes, we engaged in member checking with four Project Rebound staff members. 
This involved presenting preliminary themes and the higher-level codes that were within each theme. 
The members confirmed that the high-level codes sufficiently captured key issues and provided input in framing the themes. 
The input was incorporated in subsequent iterations where we arrived at five overarching themes, forming the basis of the findings section below.

	\section{Findings: Navigating Reentry}
	\label{sec:findings}
	In claiming that our social systems enable second chances, we assume that returning community members have sufficient support to be socially and economically integrated.
	These assumptions do not hold true for many. 
	Indeed, our field notes underscored a disconcerting observation: the system appears to be geared to re-incarcerate returning community members rather than see them make the most of their second opportunity, mirroring the sentiment expressed by \citet{Petersilia_2001} in 2001! 
	
	The participants shared several challenges they faced and continue to face in their reentry journey, and shed light on the various dimensions of support they desire.
	We also share the various manifestations of our partner organization's high-touch approach in response to those challenges. 
	The high-touch approach is centered on human connections and holds significant importance to the organization. 

	\subsection{Restrictive Programs: Supervision Over Services}
	
	Our discussions unveiled the systematic positioning of returning community members as subjects within the parole system. 
	The programs they were enrolled in often failed to adapt to their diverse needs, nor did they have a voice in shaping the programs designed for them.  
	Participants recounted their experiences in programs that they considered excessively restrictive, yet felt compelled to comply to avoid potential parole violation.

	These restrictive programs posed significant barriers to exploring possibilities, particularly during the early stages of reentry. 
	One such aspect was related to transitional housing, which are meant to provide temporary and supportive accommodation for returning community members \cite{halushka2019runaround}.
	However, these homes often operate as profit-driven entities without standardized operational protocols. 
	This leads to residents having varied and sometimes unsuitable experiences. 
	As DR highlighted, ``\textit{It's the parole officer or the parole board that make that determination the student} [returning community member] \textit{has no choice. They have to go. And then whatever modality they are sent to so often times, drug and alcohol houses, students get sent to even if they haven't had addiction issues}''.
	Participants frequently referred to transition homes as ``halfway house'' or ``prison-after-prison,'' reflecting the sense of powerlessness they experienced akin to their time in prison.

	Particularly pertinent to our work was the practice within some transition homes of completely disconnecting residents from the outer world. 
	DR shared concerns about these homes:
	\begin{quote}
		\textit{So some, not all transitional houses, but some housing modalities have what's called a blackout period ... they are prohibited from contacting anybody for 90 days sometimes ... those programs prevent students from making contact with family members or friends.}
	\end{quote}
	Further, the staff noted the challenges in accessing basic digital resources in the transition homes, ``\textit{Some transitional houses don't provide internet, they don't have a computer lab. So many students in the beginning all they have if they're lucky is a smartphone. But that's the only device that they have access to.}'' 
	
	Beyond transition homes, restrictions were also evident in parole policies. 
	We could hear this, for example, in DR's account of cases where ``\textit{Some parole officers are not big on education. They may not think that earning a college degree is a viable parole plan and they push our students to work full-time instead.}'' 
	None of the participants were confident that they could shape the parole system designed for them. 
	These restrictive programs constitute hidden yet significant barriers to reentry.

	While Project Rebound had processes in place to offer more holistic support, such as legal counseling and connecting returning community members to other services, they had limitations in effecting change within transition homes.
	Within the organization, though, they fostered a more supportive environment, one built on trust and the assurance that the returning community members would feel safe.
	As GJ, a staff shared, ``\textit{The first thing we try to do ... our best to make a person feel comfortable and to make them feel that you're no longer an environment where someone is threatening you. You are in an environment that's safe.}''
	This empathetic approach was seen as an asset by the organization.

	\subsection{Unresponsive Support Structure}
	
	
	Participants reported significant variations in their interactions with parole officers. 
	They acknowledged that the extent of resources made available to them often depended on the proactive efforts of individual parole officers, as articulated by EH, who noted, ``\textit{It's }[parole support] \textit{limited. And then some may be better than others in that they have, um, better access to resources because they make it a point to do that.}'' 
	They also highlighted that the information they received from their parole officers was frequently outdated, with many listed services (e.g., for housing) having long since closed. 
	Most participants found the services provided to be minimal, leading one participant to offer a cautionary suggestion, ``\textit{Don't depend on the system to provide everything that you need because it's not set up that way.}''
	
	A few participants navigated the lack of resources by seeking public assistance.
	EH, for example, shared how he received help, ``\textit{so most people get out, don't have money so they get on G.A. general assistance, Medi-Cal and that qualifies you for free phone.}''
	However, even when resources were available, returning community members encountered unforeseen challenges that they had no means of addressing.
	For instance, OW, in his search for financial support, discovered that he was eligible for a Pell Grant which opened the possibility of pursuing an education.
	When he applied, he was informed that ``\textit{somebody got a Pell Grant from my }[his] \textit{identity before},'' rendering him ineligible. 
	The earlier Pell Grant application had been submitted while OW was in prison and thus was a fraudulent application. 
	But there was no room for him to seek help from the parole system.
	Project Rebound eventually assisted OW, with him recounting, ``\textit{It ended up being resolved because the organization} [Project Rebound] \textit{got involved and the cost went down to the Department }[of] \textit{Education and pretty much the organization }[Project Rebound] \textit{had to pay for it.}''
	
	The unresponsiveness of the support system was also evident in the limited mental health facilities provided.
	It is well established that mental health challenges are prevalent among returning community members \cite{begun2016mental, Petersilia_2001}.
	GJ, reflecting on his experience, raised this issue, ``\textit{you don't feel what people in the greater community feel because you have to suppress so many of your feelings}'', adding, ``\textit{ therefore you're }[continuing to be]\textit{ a prisoner}''. 
	While some parole officers provided information on available mental health facilities, most did not.
	However, simply providing information is insufficient; building trust to encourage the use of the resource is critical.
	JN captured this sentiment well, sharing, ``\textit{Things like confidentiality don't exist in prison. So, this whole idea that I can go to someone and share and be vulnerable and have that information stay private is not necessarily a trusted concept.}'' 
	In this context, Project Rebound has made concerted efforts to build trust and encourage its students to utilize the university's free mental health services. They found that extensive effort is required, particularly to ensure that students trust the process and the institution.

	\subsection{Financial Struggles}
	
	Echoing previous findings (e.g., \cite{lyons2011compounded, bushway2012signaling, visher2008employment}), all participants shared their struggles with the lack of resources. 
	Our findings highlight that the challenges were most acute during their earlier days of reentry. 
	A telling example of this resource deficiency is ``gate money'', an amount defined by each state to assist returning community members with their financial needs \cite{armstrong2019what}. 
	In California, each individual is allocated \$200\footnote{\$200 is, in fact, the highest amount of gate money provided across all states in the US \cite{armstrong2019what}.}, an amount that was deemed to be sufficient to pay a month's rent in 1973 but has never been adjusted since. 
	
	The amount is insufficient to serve its intended goal. 
	JN, aptly conveyed the challenges of starting anew with just the \$200 gate money:
	\begin{quote}
		\textit{So when you get out, you get that little 200 to start your life. And usually, the first thing that people are thinking about doing is not investing in some form of computer or technology to keep them connected. It's usually about basic basic stuff like you know, set up for some meals and maybe a few outfits that will keep them, you know, so you can walk around and feel like normal. So that \$200 is not going to go too far as you build your life as far as priorities go.}
	\end{quote}
	
	Some participants, such as OW and EO, relied on friends and family for financial help. 
	Additionally, family and friends often helped them gain access to smartphones, typically providing them with older devices. 
	However, others, like GN and JT, lacked such social support. 
	Some such as JR, opted to sever ties with their past relations to ``\textit{get a fresh start}''. 
	In all these instances, the participants ultimately relied on Project Rebound to gain access to devices.
	
	Apart from the challenge of obtaining devices, accessing essential services proved to be a costly and daunting endeavor.
	For instance, returning community members relied on internet connection at our university during normal times. 
	However, when the university transitioned to online classes during the pandemic, they had to resort to drastic measures. 
	As recounted by a staff member, ``\textit{The library, the computer lab, once they have a student ID number} [you can use it] \textit{but during the pandemic, all of that was closed ... students} [returning community members] \textit{had to go to public parks to access free Wi-Fi which is not always the safest option to connect to the internet and provide your social security number and all that stuff ... oftentimes, the only internet they were able to access. The financial aspect, that they don't have a lot of money.}''
	In all these cases, financial struggles posed a significant barrier to successfully navigating our society, which increasingly assumes easy access to technology and basic services for various functions.


	\subsection{Challenges in Finding Reliable Employment}
	\label{sec:employmentChallenges}
	Financial struggles are exacerbated by challenges in finding reliable employment.
	Our participants were enrolled in formal education where they saw opportunities for employment as heard in, for example, EO's comment, ``\textit{You could take this amount of a two-year program and just get you a certificate and get your feet in the door to get a job}''.
	Even then, they encountered several multifaceted challenges during their job search. 

	First, during their preparations for the job search, participants expressed uncertainty about what to include on their resumes. 
	The staff members shared the challenges they faced when helping returning community members recognize and appreciate their existing strengths. 
	DR, for example, shared: 
	\begin{quote}
		\textit{It is about recognizing their }[returning community members] \textit{strengths. Oftentimes, they don't realize their own strengths. They don't, and we have that discussion where we talk about strengths.} [They are] \textit{like I have nothing to show for and then when we go over well what did you do while you were inside and they say `I was a welder, I did this and I did that' and that there are skills and strengths that they already have but if they gain inside, sometimes my impression is that they don't think they are as valuable because I got that in prison so how I'm gonna show that and then the fear is can I put that on my resume.}
	\end{quote}
	The fear of inadequacy or stigmatization often led returning community members to question whether they could include prison-acquired skills on their resumes.  
	Project Rebound attempted to address this issue by incorporating assets-based processes, ``\textit{around naming your own strengths and then owning your strengths, and then maybe in a job interview, talk about your strengths}.'' 
	
	Second, current job application processes assume a level of digital proficiency such as in creating resumes, writing cover letters, building portfolios, and participating in online interviews. 
	Many returning community members struggled with these digital operations and lacked reliable support systems to seek help. 
	Some got help from existing systems, but that was limited.
	For example, to create his resume, EH worked with a counselor: ``\textit{I sat with somebody who knew how to ask the right questions. They started building on that and created a resume.}'' 
	However, the help was just one session long and it involved using a resume template in Google Docs which EH was unfamiliar with, thus restricting his ability to make changes to the resume beyond that counseling session. 
	While the counselor's help is valuable, such support is far fewer for returning community members.
	On this front, Project Rebound worked to provide personalized guidance, but staff members expressed challenges in managing the workload and the technical aspects involved in the process, which JN shared, ``\textit{They want you to log every class you ever took. That used to exhaust me when I was doing that stuff but the team has gotten really good at doing it, but it's hard and people get impatient.
	}''

	Third, participants highlighted that background verification processes barred them from employment. 
	For instance, MN received a job offer but was subsequently retracted due to a background check failure. 
	The company's denial was based on a misdemeanor that occurred beyond the seven-year limit stipulated by California's legal ordinance. 
	With Project Rebound's help, MN filed a lawsuit against the company, and at the time of our study, he was waiting for the hearing date.
	
	These challenges highlight the multifaceted barriers returning community members face in pursuing reliable employment. 
	Addressing these barriers requires comprehensive support that recognizes their strengths and enhances digital literacy but it also needs broader systemic changes and advocacy for fair employment practices. 

	\subsection{Navigating the Technological Ecosystem}

	Returning community members have to navigate the complex technological ecosystem that has become an integral part of our daily life. 
	Our participants shared challenges they faced in daily tasks, such as setting up appointments with public services, opening a bank account, or taking public transit, all of which required digital know-how. 
	
	Technological vulnerability was salient in our conversations. 
	Some had fallen victim to scams. 
	For example, JD recounted an incident that occurred just a few days before our study where she lost her savings due to a scam that preyed on her empathy: 
	\begin{quote}
		\textit{It was so stupid ... So I got this email on the university email ... So I got this email about some of them having kids and that I would have to go and buy stuff for orphanages and like foster homes and all. I was a foster kid, you know, so everything like made sense to my spirit or whatever ... So finally I was like anyway} [gave them card details to pay]\textit{, and I thought it was genuine because I looked up that stuff and there was a lady with ... that baptist, on the east coast ... but like I saw something weird that happened on the bank card and I was gonna like call Chase } [bank] \textit{... because I check up my balance ... and by then it was gone}
	\end{quote}
	Considering that she had checked the organization and found it to be genuine highlights how challenging the digital ecosystem can become. 
	Such experiences deterred some participants from using technology altogether. 
	We could hear this, for example, in JT's account, ``\textit{Nah I don't trust the bank on the computer. I don't do online. I am an in-person guy ... they try to tell me banking online but it goes in one ear. I don't do that ... That's why when I took a class last year, I took an Excel class and I was kind of iffy about doing it because it was online.}''

	Participants also found it challenging to keep up with the constantly evolving technological ecosystem.
	The transition from the university's old learning management system (called iLearn) to Canvas was met with frustration and resistance.
	JT shared his frustration, ``\textit{All of technology. I just can’t like ... I was barely learning iLearn and now I am here. That is what scares me cause it is like you change the software now ... I just barely learned this thing and now I am like, okay, there is no way},'' a sentiment that other participants strongly corroborated, ``\textit{You mentioned Canvas. For me, it was like trauma!}''

	While participants wanted to learn technology, they struggled to find the right level of digital literacy programs. 
	Some programs assumed a level of technological proficiency that overwhelmed returning community members, a sentiment that EO shared, ``\textit{It is discouraging me with technology where I am like this is not for me. Because I don’t know the shortcuts. I don’t know how to organize it. I can just go here, get it from my documents, instead of having it to send it on my email and then getting it on a text message, and then have to delete it, drag it over. Oh my god! That is too much.}''
	Some relied on YouTube to learn to use technology. 
	However, they found YouTube to be overwhelming, a value that AP shared and many others echoed, ``\textit{YouTube confuses me more sometimes. Because it is step-by-step. Like I can’t follow it and get more confused. Like half the time, I am just missing the first step or something like I don’t know  I am missing one or two steps cause somewhere along the way I missed a step or something.}''
	
	As a result, they sought help from individuals rather than relying solely on digital tools, something that AP tried, ``\textit{For me now it’s like I am here in programs like this. Talking to people. This is the only way that I am going to be comfortable. Because ... you gotta learn.}''
	Indeed, the value of personalized support came up multiple times. 
	JT, for example, when talking about his refusal to online use shared, ``\textit{I'm not good online because when you get into a situation and you have nobody to talk to, you just got the wall ...}''

In some respect, our participants had more institutional support than typical returning community members.
Indeed, Project Rebound's services focused on providing one-on-one tailored help as part of their high-touch approach. 
For example, JR shared  his appreciation of a more personalized, human connection, and trust-focused help that he had received earlier through an intern at Project Rebound:
\begin{quote}
	\textit{I had one of the interns she graduated when she left to take care of} [her] \textit{grandmother, she would always reach out to me ... you know, need anything or help me support me ... so that that made me feel confident because some people do care and see you accomplished. If nobody ain't checking up on me and seeing how I'm doing, how do I know if they, you know, feel ... I look for stuff like that, the people that, you know that care, I don't want to be around nobody that don't care and I don't like a person to give me the run around.}
\end{quote}
However, staff members reported being overwhelmed at times, sharing the demand on their time for seemingly simple tasks, as DR shared, ``\textit{Sometimes it takes us half an hour for } [setting up] \textit{a password because not all of the rules are uppercase, lowercase, special character, what is the special character? ... That's high touch, right? Who has time to sit with somebody and create a login for an hour but that's what we do here.}''
\\



Overall, these findings highlight three key issues.
First, there is variance in the returning community members' journey, such as in the kind of parole support and transition home they get enrolled into, the extent of support from family and friends, and the differences in their ability to leverage digital technologies. 
Second, there is a significant gap between the needs and the availability of support for the returning community members; the lack of responsive support is quite striking.  
Third, our partner organization's high-touch approach provided some of the needed support, which, while effective, has limitations due to a lack of structure and an inability to scale.

\section{Implications to Our Design}
\label{Sec:implications}

\begin{figure}
	
	\centering
	\includegraphics[width=1\linewidth]{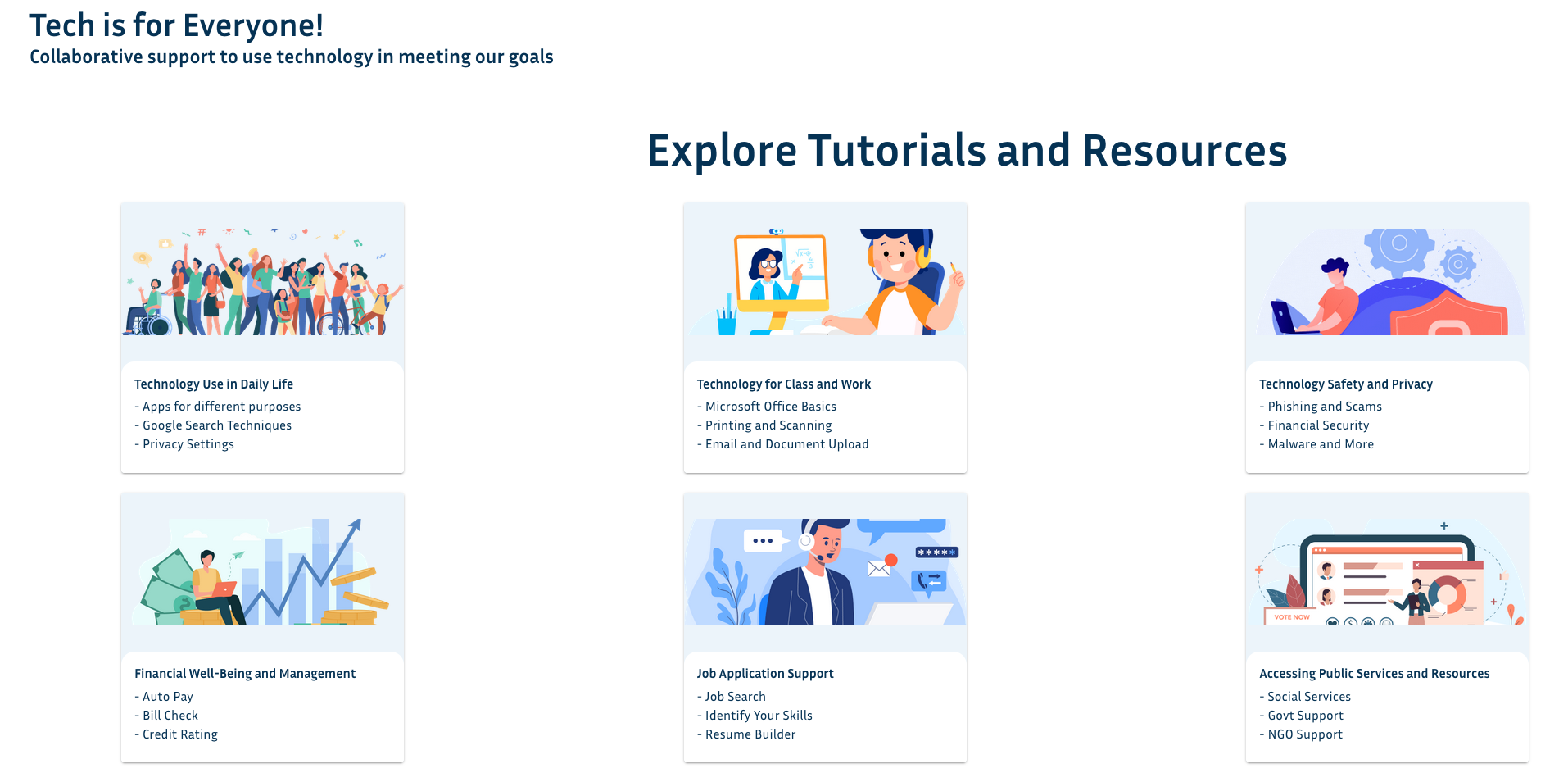}
	\caption{The web application's landing page displays tutorials and resources grouped into six categories.}
	\label{fig:categories}
	\Description{The image shows our web application's landing page. The six categories help are arranged in six cards, with three in each row. Each card has an image on the top with a title followed by a list of sub-titles describing the topic of help included within the card. The upper row covers three categories, namely Technology Use in Daily Life, Technology Use for Class and Word, Technology Safety and Privacy. The lower row has three cards for the rest of the three categories, which are Financial Well-Being and Management, Job Application Support, and Accessing Public Services and Resources.}
	
\end{figure}

The findings from our initial study had several implications for our next steps.
As we heard in the findings, supporting technology use among returning community members was a major concern for Project Rebound. 
To this end, they were exploring the possibility of creating tutorial videos, such as to setup passwords.
They envisioned using the videos to augment their personalized high-touch approach, using it to offload some of the common technical aspects while the staff provided more personalized help when needed.
This priority prompted our exploration of potential technological solutions using existing videos.

We recognized the necessity to integrate the technology within Project Rebound's existing processes.
The human connection Project Rebound built with the returning community members was significant and helpful to the returning community members.
The returning community members trusted Project Rebound; a standalone tool would not accord similar trust. 
Also, for long-term sustainability, creating a tool for the organization would more likely lead to relevant updates and continued use, even after our project ends. 

Incorporating the tool within the organization's processes entailed involving the staff members and returning community members throughout the design process (Section \ref{sec:researchApproach}). 
While this constitutes a classic human-centered design approach \cite{lazar2017research}, we tried to build their sense of ownership over the tool, clarifying that we want to hand over the tool and that they should make decisions accordingly. 
The tool, a web application, was named after the organization, reinforcing the notion of organizational ownership and building on the concept that trust in the organization can extend to trust in the tool \cite{mayer1995integrative}.
We share the two key principles that guided our design of the web application: incorporating it into the organization's existing work and designing to attend to the variations in technical skills.

\subsection{Integrated Design: Incorporating Into the Organization's Existing Work}
Thematic analysis of the problems mentioned in the initial inquiry led us to identify six categories surrounding digital skills and literacy: Technology Use in Daily Life, Technology for Class and Work, Technology Safety and Privacy, Financial Well-Being and Management, Job Application Support, and Accessing Public Services and Resources. 
The web application's homepage prominently featured resources categorized under these six headings (see Figure \ref{fig:categories}).

\subsubsection{Staff Members Control the Content} 

All sections except Job Application Support and Accessing Public Services and Resources contained videos.
We populated these sections with select videos from YouTube, specifically chosen to address the challenges participants had mentioned with technology (e.g., inserting an image into a Word file).
These videos were initially populated with the goal that the staff members would curate their own set of tutorials as needed. 
They had, in fact, already started exploring the possibility of creating tailored videos and uploading them to YouTube.  

Our web application supported the staff members to upload videos. 
Uploading a video involved entering the video URL, tags, selecting the device(s) for which the video is intended, specifying the category, and optionally including time markers and corresponding message for that segment. 


\paragraph{Feedback:} The aspect of content control was well-received by staff members. 
They found the video upload process to be straightforward and efficient, with MN remarking, ``\textit{I've uploaded videos to YouTube, and it's like a whole process. But this is just super simple, you know? You scope the videos, and then you just add it on there, and then it uploads super quick.}'' 
There were some concerns about not being able to add videos locally, as DR shared, ``\textit{So we should be able to add those here, right? But what if it's not a YouTube video} [an external URL]?''  
We clarified that we wanted to be frugal so we relied on other locations to host the videos (e.g., YouTube or Google Drive) and did not support administrators to host videos locally. 
The clarification helped as was shared by another staff member, ``\textit{It's pretty much straightforward. It's not too confusing with too much fluff, you know. I like it.}''

\subsubsection{Job Application Support}

A significant part of the organization's work involves helping returning community members prepare for job searches. 
Our findings highlighted multifaceted barriers faced in job searches that Project Rebound staff members helped with (see Section \ref{sec:employmentChallenges}). 
Considering these, we designed a set of features in the web application to support the staff members.

\begin{figure}
	\centering
	\includegraphics[width=1\linewidth]{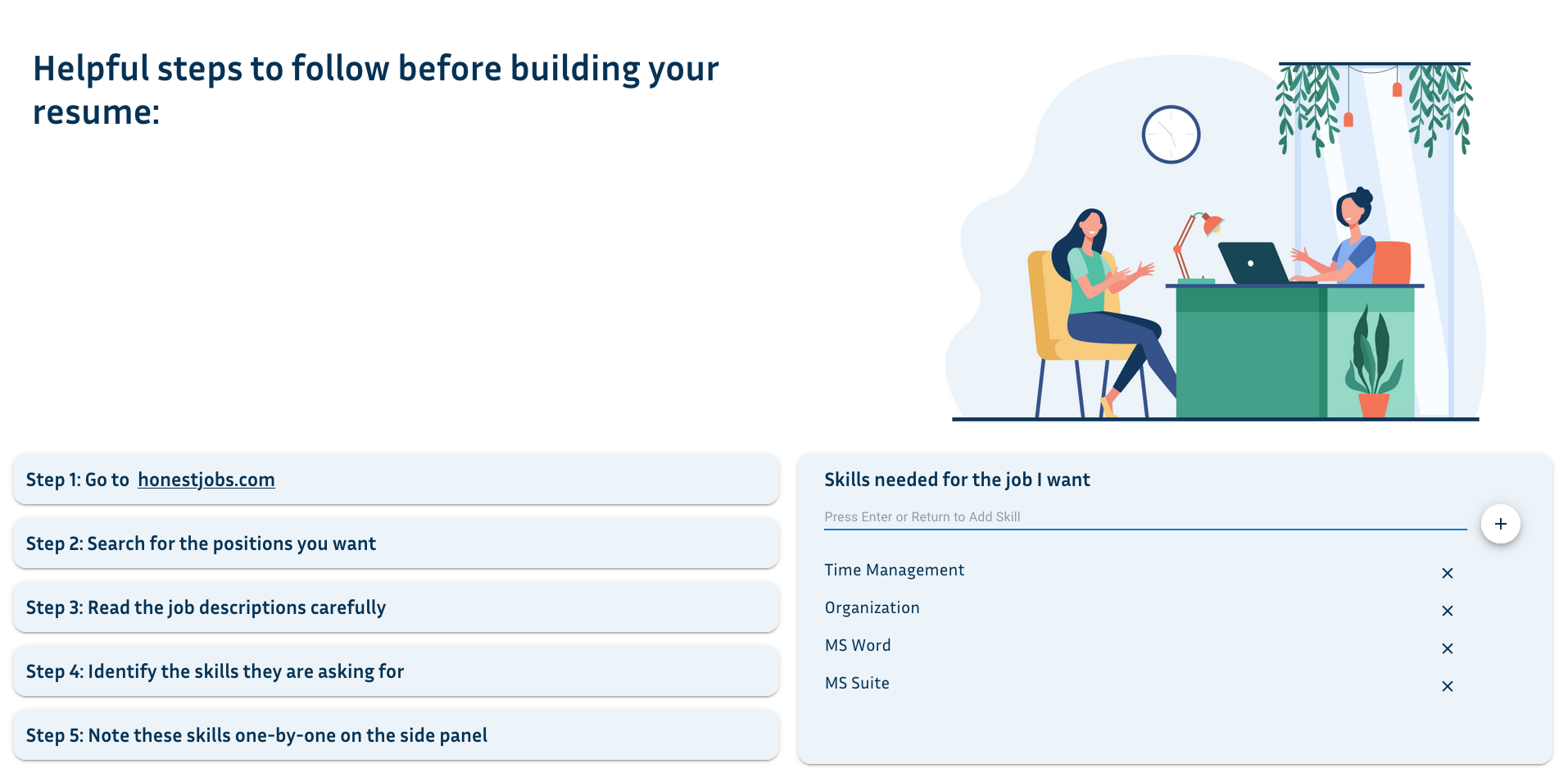}
	\caption{Guided job search feature asks users to search for aspired jobs and list their skills to highlight in their resume.}
	\label{fig:jobSupport}
	\Description{This image shows the job search page. It contains an icon on the top along with text that reads ``Helpful steps to follow before building your resume:'' Below it, the image lists five steps asking users to go to \url{honestjobs.com}, search for aspired job positions, read those job descriptions, identify skills listed on those positions, and note them down. Towards the right of the image, there is a small note-taking interface that allows users to note down the identified skills.}
\end{figure}

\begin{figure}
	\centering
	\includegraphics[width=1\linewidth]{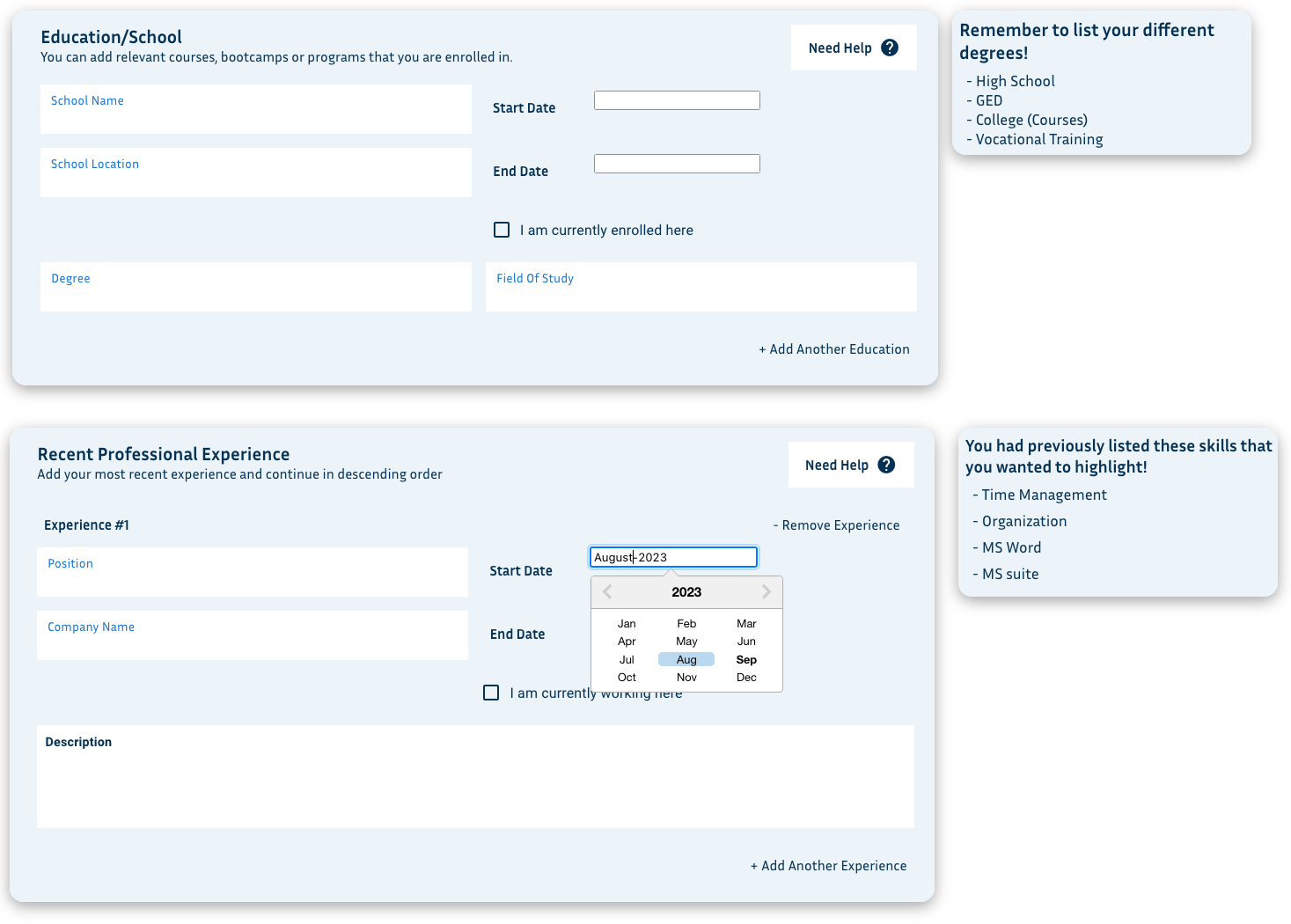}
	\caption{Resume builder form with key reminders on the side to scaffold reflection and enable collaboration with staff members.}
	\label{fig:resumeReminder}
	\Description{This image shows two major segments of the resume builder form. The top segment is about education and shows a reminder box that reminds users of different degrees they may have accomplished. The lower segment is about recent professional experience which also has another reminder on the right side which lists out all the skills users had noted to highlight earlier.}
\end{figure}

First, staff members noted that most returning community members were reflective but needed help in identifying their existing strengths.  
We designed a feature that guides users to search job listings, identify the key skills required for desired positions, and document those skills they already possess (Figure \ref{fig:jobSupport}). 
Users are later reminded of these skills when building their resumes in our system.
The resume reminder served to facilitate self-reflection while also enabling the staff members to encourage returning community members to include those overlooked skills that they possess. 
Similarly, in the education section, we included a reminder on the side encouraging users to detail all their accomplishments, extending beyond formal education degrees (Figure \ref{fig:resumeReminder}). 

Second, returning community members faced challenges in creating and maintaining resumes. 
Resume building requires detailed knowledge of formatting, versioning, and working with various file formats (e.g., Word or Google Docs, and PDF).
Staff members also struggled to keep track of multiple versions and formats when providing one-on-one resume guidance.
To help with this, we created a resume builder that enabled the staff members and returning community members to focus on content creation while our system facilitated formatting and version control (Figure \ref{fig:resumeReminder}). 
Users can fill out a form and, when needed, generate a formatted resume in PDF that could be parsed correctly by automated resume filtering systems.

Third, the staff members, as part of their high-touch approach, provided in-person troubleshooting for various digital challenges. 
We sought to embed this ethos by including ``Need Help'' buttons on major elements of the web application. 
Each section of the resume builder, for instance, featured this button (see Figure \ref{fig:resumeReminder}). 
The Help Manager page enabled staff members to track users' requests, fostering a collaborative environment wherein returning community members could work independently, reassured by the knowledge that assistance was readily available when needed.

Fourth, during an early iteration, we asked returning community members to use Indeed\footnote{\url{https://www.indeed.com/}}, a popular job board. 
Staff members shared that most returning community members are unable to start their jobs even after successfully passing the interviews, as they fail their background verification. 
While this violates the essence of the state law (California's Senate Bill 731 and Assembly Bill 1076), companies have devised legal loopholes to reject candidates without violating the letter of the law. 
One staff member, JN, even suggested the need for a community-curated list of companies that hired individuals with criminal histories, likening it to a modern-day ``Green Book''\footnote{Referring to Victor Hugo Green's guidebook to locations and services that were friendly to Black Americans during the Jim Crow laws era.}.
A similar problem was identified by \citet{Dillahunt_2016} who also saw a need for curating a list of ``felony friendly companies'' when designing systems for disadvantaged job seekers. 
MN shared that he used Honest Jobs\footnote{\url{https://www.honestjobs.com/}}, a job board that curates lists of companies known to hire people with a criminal record. 
Our guided job search process now features HonestJobs (see Step 1 in Figure \ref{fig:jobSupport}).

\paragraph{Feedback:} The job search and resume builder tool garnered enthusiastic responses from the returning community members. 
For instance, AP found it particularly beneficial for individuals with limited digital skills, ``\textit{This right here to me is like basic entry-level and there’s a few people who come home from prison and are like really advanced in technology but the majority this is like the entry-level they need to start on.}'' 
AP also expressed appreciation for the key skill reminders on the side panel, noting, ``\textit{The highlight is giving you the language and when they are asking you things like time management, you will be put in like yes you look at the clock, but you are giving the language} [to write it].'' 
EH, who had previously encountered challenges with resume templates, shared his positive sentiments, ``\textit{That is a brilliant idea. I wish I had that when I came home.}''

\subsection{Low-Floor High-Ceiling Design: Attending to Differing Levels of Technical Skills}
\label{sec:lowFloor}
Both the staff and participants acknowledged the wide differences in digital skills, with one staff member noting, ``\textit{I think that} [digital skills]\textit{ correlates a lot with how much time a student has served in prison right? I know students who have maybe done five or eight years they were on the outside when the iPhone was invented. If I talk to a student who has been in prison for 23-27, 30 years they usually type with one finger. They don't know how to turn the computer on and off, where the on and off button is.}''
To accommodate this variation, we designed our system following the low-floor high-ceiling principle \cite{resnick2008falling}, which aims to provide an easy entry point for all users while offering opportunities for all.

\subsubsection{Easy Entry and Access}

All materials and resources, except for the resume builder, are accessible without requiring users to log in. 
Users only need to sign up or log in if they wish to use the resume builder, as we save the information entered in the form.

Users also have the option to specify the mobile and laptop devices they have, either while signing up or later in their profile settings.
This information is used to filter video tutorials in each category to match the user's device type. 
This feature aims to reduce the overwhelming number of videos presented to users, given the different device types and usage scenarios. 
In our subsequent iteration, we also incorporated a set of 12 questions about laptop and mobile use to recommend a set of videos for learning critical digital skills.

\subsubsection{Supporting Access to Public Resources}

The staff and returning community members shared difficulties arising from limited resources.
Outdated information about potential resources was also a common challenge.
Digital literacy includes knowing how to access information, such as accessing public resources and infrastructures.
To help with this, we integrated Findhelp\footnote{\url{https://www.findhelp.org/}}, a local resource registry that has dedicated database managers to verify and update support programs\footnote{We are deeply indebted to Ihudiya Finda Ogbonnaya-Ogburu for sharing this resource with us.}. 
The service is increasingly being used by social work and health care scholars (e.g., \cite{shadowen2022exploring, dimick2022addressing}). 
We found it to be reliable and comprehensive.
We incorporated the service within our web application with the plan that this would help the returning community members learn to leverage an existing tool. 
The staff members recognized the value of this external service, as it could continue to assist returning community members even after their involvement with Project Rebound.

\subsubsection{Structuring Video Tutorials}
A key issue shared by the returning community members was that they found YouTube tutorials to be overwhelming.
YouTube has plenty of resources for people of varying digital skills, but they are rarely curated or sufficiently structured to support learning. 

Considering this, we designed an approach to structure the interaction with YouTube videos. 
Staff members have the option to create segments and add prompting questions to confirm the user's understanding of the covered content when uploading a video.  
When a user plays a video and arrives at the end of a segment, the video will pause, prompting a dialog box with three options: ``Restart from the beginning,'' ``Restart from the last segment,'' and ``Doing ok. Carry on.'' 

\paragraph{Feedback:} The participants appreciated the interactive feature, ``\textit{I think it looked really well. It was again pretty straightforward in terms of like Restart, go back to the last segment, ok, continue. I think that was really helpful ... when I don't understand something I'm looking at or a tutorial, I always need to go back myself, and I need to like find exactly where I need to go if I didn't get it right the first time.}''

\section{Discussion}

Our findings corroborate prior research highlighting the multifaceted challenges that returning community members encounter during their reentry journey \cite{williams2019s, baer2006understanding, frazier2014african, travis2005but}. 
We also noted significant variations in the reentry experiences. 
These were shaped by factors such as participation in transition programs, support from family and friends, educational background, technical skills, and the duration spent both inside and outside prison walls.

We observed our partner organization's various efforts to support returning community members.
Their high-touch approach, emphasizing building connections and providing personalized guidance, was prominent.
It helped address the many gaps in support available to returning community members during their reentry journey.
The connections built with the staff, for example, when searching and applying for jobs, served to moderate the absence of encouragement and support available elsewhere.
The substantial workload borne by staff members in implementing this high-touch approach compensated for the lack of material resources necessary to provide such tailored reentry support. 
This raises questions about its long-term sustainability and effectiveness if it needs to be scaled. 

In the face of structural problems confronting returning community members, it is evident that technology, while valuable, is not the solution.
It can be a supplementary resource to the critical efforts of partner organizations and their support networks.
Our design and engagement are firmly rooted in this belief, seeking to enhance the partner organization's existing efforts, specifically in promoting digital literacy.

It is crucial to recognize that our participants do not represent the typical population of returning community members in some regard.
Most were either trying to enroll or were already enrolled in formal education programs at a public university and had access to institutional resources.
These resources are not readily available to the majority of returning community members \cite{halushka2019runaround, ortiz2019system}. 
These assets shaped the returning community members' perspectives and possibilities for reentry, which, in turn, shaped our approach.
For example, our designed system assumes that the jobs returning community members aspire for will be listed on Honest Jobs. 
This was true for our participants. 
However, these are not the only kinds of employment opportunities available, nor do all returning community members aspire to such jobs. In those cases, using our system may not be helpful. 
It could, in fact, enforce a narrow pathway for reentry and cause harm. 
Despite such advantages, our participants continued to face significant hurdles, which further highlight the enduring challenges in reentry.

\subsection{Implications for Designing Digital Literacy Support}

Given the significant differences between our participants and the larger population of returning community members, our findings may not be generalizable. 
However, some of what we learned can be helpful to other HCI practitioners and researchers, particularly those seeking to promote digital literacy among marginalized groups. 
We reflect on our experience to draw out the following four lessons.

\subsubsection{Supporting Continuity of Care}

Digital literacy is not solely about gaining information isolated from the realities of the surrounding context. 
Digital literacy entails leveraging communication technologies to achieve goals that matter to the individual.
Thus, to support digital literacy, we need to engage in socially situated practices that go beyond the digital environment. 
We argue that supporting digital literacy requires enacting care in its broadest sense, as the ``activity that includes everything that we do to maintain, continue, and repair our `world' so that we can live in it as well as possible'' \cite[pp. 34]{fisher1990toward}.


Our partner organization's high-touch efforts embody such continuity of care which extended beyond learning about technology, as we saw in their efforts against unfair hiring policies or rectifying an identity theft case. 
We also saw how returning community members encounter discontinuities in care, such as being assigned transition homes that do not meet their needs or abruptly losing access to devices and internet in transition homes. 
At the same time, we noted the personalized care from the staff members helped bridge some of the gaps arising from unresponsive and limited reentry services. 
The support demanded the staff to go well beyond their prescribed roles and responsibilities, and we heard accounts of exhaustion. 

Our design responded to the importance of care by building upon the partner organization's high-touch approach. 
It involved creating a space where the staff and the returning community members could collaborate and explore together, promoting interdependence rather than co-dependence.
For example, the resume builder encouraged working along with the staff members, and the ``Need help'' feature within each element of the resume builder provided additional space for the staff to enact care when returning community members worked alone.
At the same time, the design sought to reduce the load on the staff \emph{without} replacing their care practices. 
The technology was designed to be integrated into the existing practices.
For example, the reminders of existing skills and auto-formatting of the resume sought to help the staff members as they supported the returning community members. 


In contrast, contemporary technology design, in the pursuit of scalability and efficiency, has relied on virtual and remote support mechanisms, assuming a certain level of digital proficiency \cite{biehl2022can, van2014digital}. 
Critically, they are devoid of care practices. 
This deficiency was evident, for example, when participants expressed being overwhelmed with tutorials on YouTube or refused to use technology when they felt there was no room for seeking help. 
This leads us to question the use of scalability and efficiency as metrics for designing and evaluating digital literacy systems.
Critically, there is a risk in automating learning especially when returning community members desire human connections and a sense of belonging.
In this respect, we echo \citet{tseng2022care} and advocate for prioritizing infrastructures that enable continuity of care, building upon existing care practices. 

\subsubsection{Promoting Reflection Through Slow Technology}

We advocate for designing slow technologies to support digital literacy.
Slow technology is a design philosophy that argues for ``exposing technology in a way that encourages people to reflect and think about it'' \cite[pp. 204]{hallnas2001slow}. 
By inviting users to develop a relationship with technology over time, slow technology facilitates a deeper connection with oneself and one's surroundings, enabling individuals to recognize and build upon their own capabilities.

In an era where there is a growing push for technologies that generate artifacts rapidly, such as using large language models to build resumes (e.g., \cite{chatgptResume}), we call for paying attention to (and designing) slow technology.
In proposing slow technology, \citet{hallnas2001slow} assert that it can provide opportunities to learn how technology works and why it does so. 
Our findings support the argument.
In our initial inquiry, the staff member shared that returning community members often under-appreciate their existing strengths. 
In our design, we asked returning community members to go to Honest Jobs and identify key skills to highlight in their resumes. It helped create a space to reflect on and appreciate their accomplishments. 
Participants also shared that this exercise helped them learn new keywords relevant to their desired job positions.

Slow technology design prioritizes slowness in its appearance and presence, as well as in its use of clear and simple material \cite{hallnas2001slow, odom2012photobox}.
These principles are important when designing for (digital) literacy. 
Slowness in appearance and presence allows individuals the necessary time to reflect, build skills, and nurture relationships, both with technology as well as with other actors involved in the process.
The use of clear and simple materials lowers barriers to entry, reduces novelty effect, and creates space for people with diverse skills to engage and learn by drawing upon their existing strengths and assets. 

Further, slow technology can draw attention to the aesthetics of functionality, prompting critical reflection on technology's various functions and implications.  
For example, PhotoBox's slow design draws attention to the rapid and massive consumption of photos enabled by digital cameras \cite{odom2012photobox}.  
In the case of digital literacy, slow technology could facilitate reflection on various functions of technology such as automatically filtering job applications or conducting background checks.
Such critical reflection is at the core of digital literacy. 
While our current work has not fully explored this aspect, we recognize it as a significant opportunity for future research in designing digital literacy support that empowers individuals to critically engage with technology.

\subsubsection{Building in Flexibility}

Returning community members are not a homogeneous group.  
As our findings highlight, variations extend beyond digital skills; they manifest in multiple facets of reentry, such as parole restrictions, level of family support, and personal aspirations and goals. 
These variations can influence the adaptation and impact of digital literacy approaches.
Recognizing the diverse strengths, assets, and needs of returning community members, we argue for incorporating flexibility as a core principle in supporting digital literacy.

Low-floor, high-ceiling design principle guided our efforts to support individuals with varying digital skills, making it easy to get started for those with lower digital skills while offering opportunities for those with more skills to gain something from their explorations (see Section \ref{sec:lowFloor}). 
Other features like device-specific video curation also accommodated differential access.

Flexibility also entails attending to the individual and socio-economic realities surrounding the reentry journey.
For instance, literacy modules and services need to align with the user's broader goals and aspirations \cite{Ogbonnaya-Ogburu_2019}.
In our work, we introduced technology within contexts such as financial well-being. It helped present technology as an enabling means to other ends that they value, thereby accommodating diverse user needs and aspirations.
Similarly, recognizing our participants' interest in learning technology to excel in classrooms led us to curate a separate category of tutorials called ``Technology for Class and Work''. 
This enabled access to resources for those attending classes and aspiring to work in technology-focused jobs -- a majority of our participants -- while allowing others to ignore it if it did not align with their goals.

Another dimension to support flexibility involves ensuring learning opportunities across different modalities and contexts. 
We heard of the challenges related to accessing technology in transition homes, where many returning community members lacked access to laptops.
Some of our participants struggled to move between smartphones and laptops. 
These call for carefully considering how the designed digital literacy technology can be accessed, including the possibility of building unplugged lessons \cite{bell1998computer}. 
In our case, our web application was primarily designed for use within the organization, supporting the staff member's interactions with returning community members. 
The web application was responsive, enabling use through smartphones within and outside the organization, such as in transition homes where smartphones were permitted.
However, a limitation of our work, on this front, is that it still relies on Internet connectivity and device access; we are exploring the possibility of creating progressive web apps and paper-based guidebooks to reduce reliance on the Internet. 

Building in flexibility enables accounting for the variance and complexity of the socio-economic realities that shape returning community members' reentry journey. 
It also opens up space for us to acknowledge and attend to their dynamic needs as they navigate reentry within the rapidly evolving technological landscape.

\subsubsection{Reconfiguring Toward Existing Infrastructure}

Literacy is infrastructural \cite{disessa2000changing, vee2017coding}.
It enables ways of knowing and doing things.  
A key element in promoting (digital) literacy, therefore, involves supporting individuals to leverage their literacy skills to make the most of available infrastructure. 
We need to reconfigure our design to support engagement and relationship-building with \emph{existing} infrastructure. 
This contrasts with the common approach of introducing novel interfaces and systems for people to learn. 

For example, in our interviews, participants shared their struggles stemming from limited access to relevant and current information, particularly regarding resources and services tailored to their specific needs.
To help with this, we integrated Findhelp, an external system that provides an online registry of local resources, rather than creating an independent catalog of services. 

We draw attention to several considerations when reconfiguring design to engage with existing infrastructure.
If not carefully vetted, some systems may be exploitative. 
External systems can also fall into disrepair or ``die'' \cite{corry2022does}.
Hence, it is critical to evaluate the system(s) carefully, particularly by analyzing their history of stability and maintenance, and, where available, their communication of changes. 
We also need to ensure that our users have sufficient support in case the system fails; 211 calls were an alternative that we considered if Findhelp failed. 
However, in our evaluation, we found Findhelp to be reliable and well-maintained. 
Additionally, the presence of dedicated staff verifying and updating information makes it a valuable resource, such resources are inaccessible to us academic researchers and a non-profit organization.

By incorporating existing infrastructure into our design, we seek to support continuity and long-term transformation, enabling returning community members to build upon their strengths and assets to access resources and opportunities that extend beyond the immediate scope of our intervention.
In our case, the use of Findhelp opens one such possibility for returning community members to access needed resources even beyond their time with our partner organization.
We contend that reconfiguration encourages us to transcend superficial novelty, prompting us to think critically about the broader ecosystem in which our design operates.



\subsection{Reflecting on Being a Partner on the Side}

In this work, we take the role of ``partner on the side'',  collaborating with CollegeBound, an organization embedded in a network offering comprehensive reentry support services.
As HCI researchers increasingly embrace community-based approaches, particularly in engaging with marginalized populations (e.g., \cite{liang2021embracing, harrington2019deconstructing, hui2020community, wong2021reflections}), we assert that being a partner on the side is a valid research position. 
This role allows us to leverage our technical expertise in supporting organizations already addressing complex societal issues.
It is a move to reconfigure design relations around continuity and community self-determination, shifting the focus away from academic goals.

HCI researchers, including us, have often positioned ourselves as ``friendly outsiders'' in community engagement \cite{greenwood2006introduction, hayes2014knowing}.
\citet{greenwood2006introduction} liken being a friendly outsider to coaching. 
They expound the responsibilities of a friendly outsider, including bringing an outsider perspective, connecting communities with external resources, and taking an assets-focused approach by supporting local people to reflect on their tacit knowledge and assess resources that can help enact the change they desire \cite{greenwood2006introduction}. 
We strongly believe that these values are essential in community-based design efforts. 
However, positioning ourselves as a coach-like entity simplifies the entanglements that arise in such collaboration; we ought to have the proverbial ``skin in the game''. 
In particular, when examination or design of technology is involved, our technical expertise entails taking responsibility for the technology and its intended, as well as unanticipated, consequences. 
As Leigh Star puts it, ``we involve ourselves in many potential actions; these become meaningful in the light of collective consequences, jointly negotiated'' \cite[pp. 50]{star1990power}.
Partner on the side acknowledges the collective consequences, that our engagement shapes the context and we are shaped by it. 
Thus, a successful collaboration does not allow us to \emph{remain} friendly outsiders.

We note that we \emph{were} outsiders and, indeed, becoming a partner on the side requires humility and recognition that we have a limited understanding of the realities on the ground. 
As partners on the side, we need to invest time in understanding the realities faced by the community (as briefly described in Section \ref{Sec:Background}) before initiating research.
This ``work before the work'' \cite{wong2021reflections} is needed to build trust and adapt our approach attending to the historical and social context of the setting, which is critical for an equitable community-based engagement \cite{harrington2019deconstructing}.

Importantly, being a partner on the side encourages us to stand with communities as they determine paths forward based on historical and social context as well as their strengths and priorities. 
The communities are not asked to act on outsiders' terms. 
Particularly in academia-community collaborations, such as ours, being a partner on the side draws attention to de-center academic goals during community engagements. 
Community collaborations do not, and should not, revolve around academic cycles \cite{harrington2019deconstructing, pierre2021getting}.
Neither should the knowledge and recognition gained exclusively benefit the research team \cite{pierre2021getting, cooper2022systematic}, nor should it place affective and epistemic burdens on the community \cite{pierre2021getting, dourish2020being}. 
Considering these issues, community-academia collaboration benefits from community-driven initiatives. 
Being a partner on the side entails lending support to such initiatives, augmenting the community's existing capacities and assets. 
The outcomes of such collaboration shape or integrate current workflows rather than introducing wholly new systems.
It also provides space to identify, appreciate, and amplify diverse forms of knowledge production and social action, beyond those typically modeled in academic research.

In design research, positioning ourselves as partners on the side calls attention to resisting the impulse of technological solutionism \cite{pierre2021getting}. 
Social problems require social solutions; they cannot be solved with technological solutions. 
Technology can be \emph{an} element in the social solution.  
Being a partner on the side entails contributing technological and other resources \textit{as supplementary elements} to the response driven by the communities. 
In doing so, we de-prioritize technology-first solutions. 
While in our case technology was designed, it was in service of the partner organization's existing initiative of supporting digital literacy.
Equally importantly, being a partner on the side confers responsibility for the technology, including ensuring new dependencies do not arise from it. For instance, maintainability and sustainability are critical considerations \cite{kruger2021takes, hayes2014knowing, meurer2018designing}. 

Critically examining our engagement as partners on the side, we acknowledge that we should be doing more. We plan to. For example, we are writing a grant proposal together building upon this initial work.
Other future work focuses on building community capacity and supporting community ownership of the technological artifact.

In sum, being a partner on the side requires us to embrace interdependence in community engagement \cite{escobar2018designs}.
It positions us to embrace our responsibilities in realizing a fairer and equitable collaboration, going beyond the demarcations of insider-outsider status \cite{liang2021embracing} and dive deeper into formulating community-led solutions to complex social problems. 



\section{Conclusion}

Through our study, we uncovered structural barriers faced by returning community members (formerly incarcerated individuals), including support programs that prioritize supervision over service, unresponsive support systems, limited access to resources, financial struggles exacerbated by restricted employment opportunities, and technological barriers.  
We must make a concerted effort to address such structural barriers to achieve a society that truly provides second chances to returning community members. 
In our work, we collaborated with a non-profit organization that provides holistic support to returning community members, helping them overcome some of those structural barriers. 
We play a role as a partner on the side, facilitating ongoing work. 
To that end, our incremental work led us to design a web application that integrates into the organization's existing processes and supports the organization's efforts to promote digital skills among returning community members. 

We are left with several questions and areas for further examination, such as how we can support critical digital literacy where returning community members are not just learning a few digital skills but are empowered to leverage those digital skills to access resources and take actions to enact change that they desire. 
Furthermore, we have yet to address how we can support the organization and the returning community members in keeping up with the rapidly changing digital ecosystem. 
There are also concerns about the scalability of the digital system, which is dependent on the organization's valiant efforts.
Critically, perspectives from other stakeholders who are part of the reentry support system, such as family and friends, potential employers, and other organizations, would enable the design of a more comprehensive approach.  
In these respects, our work only scratches the surface, but in doing so, we hope to invite other interested HCI researchers to join us in realizing a more inclusive and supportive future for returning community members.

\begin{acks}
	We are deeply thankful to the staff members of Project Rebound and the returning community members who participated in the study and guided us throughout.   
\end{acks}

\bibliographystyle{ACM-Reference-Format}
\bibliography{dis24-56}

\end{document}